# Plasmonic Luneburg and Eaton Lenses

*Thomas Zentgraf [1,\*], Yongmin Liu [1,\*], Maiken H. Mikkelsen [1,\*], Jason Valentine [1,3], Xiang Zhang [1,2,†]*

[1]NSF Nanoscale Science and Engineering Center (NSEC), 3112 Etcheverry Hall,

University of California, Berkeley, CA 94720, USA

[2]Materials Science Division, Lawrence Berkeley National Laboratory,

1 Cyclotron Road, Berkeley, CA 94720, USA

[3]Department of Mechanical Engineering, Vanderbilt University,

Nashville, TN 37235, USA

*These authors contributed equally to this work.

†To whom correspondence should be addressed. E-mail: xiang@berkeley.edu

**Plasmonics is an interdisciplinary field focusing on the unique properties of both localized and propagating surface plasmon polaritons (SPPs) - quasiparticles in which photons are coupled to the quasi-free electrons of metals. In particular, it allows for confining light in dimensions smaller than the wavelength of photons in free space, and makes it possible to match the different length scales associated with photonics and electronics in a single nanoscale device [1]. Broad applications of plasmonics have been realized including biological sensing [2], sub-diffraction-limit imaging, focusing and lithography [3-5], and nano optical circuitry [6-10]. Plasmonics-based optical elements such as waveguides, lenses, beam splitters and reflectors have been implemented by structuring metal surfaces [7, 8, 11, 12] or placing dielectric structures on metals [6, 13-15], aiming to manipulate the two-dimensional surface plasmon waves. However, the abrupt discontinuities in the material properties or geometries of these elements lead to increased scattering of SPPs, which significantly reduces the efficiency of these components. Transformation**

**optics provides an unprecedented approach to route light at will by spatially varying the optical properties of a material [16,17]. Here, motivated by this approach, we use grey-scale lithography to adiabatically tailor the topology of a dielectric layer adjacent to a metal surface to demonstrate a plasmonic Luneburg lens that can focus SPPs. We also realize a plasmonic Eaton lens that can bend SPPs. Since the optical properties are changed gradually rather than abruptly in these lenses, losses due to scattering can be significantly reduced in comparison with previously reported plasmonic elements.**

Transformation optics as a general design methodology realizes a desired optical path and functionality by spatially varying the optical material properties, which is in contrast to the traditional methods of shaping the surface curvature of objects (e.g. lenses) to refract light. Early transformation optics devices, such as the invisibility cloak [18], generally required spatial variation of anisotropic materials and extreme values for both permittivity and permeability. Although such demanding material properties can be implemented by metamaterials, the narrow bandwidth and high loss normally associated with metamaterials limits the functionalities of the devices. Therefore, there have been great efforts in realizing new transformation optics devices based on non-resonant and isotropic materials with spatially varying properties. Examples include the carpet cloak [19, 20], the photonic black hole [21], the optical "Janus" device [22], and the flattened Luneburg lens [23]. The realization of such devices and elements is essentially based on gradient index (GRIN) optics. Compared with classical lenses, GRIN lenses exhibit the advantages that they can be flat and free of geometrical aberrations. In fact, based on variable refractive index structures even more sophisticated elements like the Maxwell fish-eye lens, the Luneburg lens and the Eaton lens were proposed more than half a century ago [24,25] but have not been realized in three dimensional optics to date.

Recently, it has been proposed to apply transformation optics to plasmonic systems, aiming to manipulate the propagation of SPPs in a prescribed manner [26-29]. If we rigorously follow the transformation optics approach, both the metal and dielectric materials have to undergo a coordinate transformation to modify the propagation of SPPs. However, since most SPP energy resides in the dielectric medium at frequencies apart from the surface plasmon frequency, it was proposed that only transforming the dielectric medium is sufficient to mold the propagation of SPPs [26, 27]. Furthermore, the transformed dielectric materials can be isotropic and non-magnetic if a prudent transformation scheme is applied. Instead of directly modifying the permittivity of the dielectric medium, we have proposed to slowly change the thickness of an isotropic dielectric cladding layer, and hence the local effective index of SPPs [27]. In such a way, the propagation of SPPs can be controlled without directly modifying the metal surface or adding discrete scattering structures on the metal. As the local effective index of SPPs is varied gradually in a truly continuous manner we term our approach GRIN plasmonics, in analogy to the well-known GRIN optics.

In this work, we demonstrate a plasmonic Luneburg lens and a plasmonic Eaton lens as a proof of principle of GRIN plasmonics. Both lenses require a gradual change of the mode index which in general is difficult to obtain in traditional optical elements. The plasmonic Luneburg lens (Fig. 1a) is similar to a traditional Luneburg lens [24] in that it focuses SPPs to a point on the perimeter of the lens. The concept of the Luneburg lens was later generalized by Eaton [25] for spherical lenses. As shown later, such lenses can lead to beam deflection [32]. Here, we realize both structures for the propagation of SPPs by spatially varying the height of a thin dielectric ($\varepsilon$ = 2.19) Poly(methyl methacrylate) (PMMA) film on top of a gold surface.

The index distribution of a traditional Luneburg lens satisfies

$$n(r) = \sqrt{2 - (r/R)^2} \qquad (1)$$

where $R$ is the radius of the lens and $r$ the distance to the centre. To implement a plasmonic version of the Luneburg lens, the effective mode index of the SPPs should spatially vary according to Eq. (1). For a lens diameter of 13 µm this leads to the mode index profile shown in Fig. 1b. Changing the height of a dielectric cladding layer is a simple way to modify the SPPs effective index [27]. At a wavelength of 810 nm, the effective mode index of SPPs on a gold film can be changed from 1.02 to 1.54 as the height of the PMMA increases from zero to 500 nm (see Fig. 1c and Methods section). Once the relationship between the SPP effective index and the PMMA height is known, the height profile for the Luneburg lens can be readily interpolated to satisfy Eq. (1). Based on the local height profile of the PMMA structure measured by atomic force microscopy (AFM), we perform an additional three-dimensional full wave simulation (COMSOL Multiphysics) to verify the theoretical performance of the Luneburg lens. The result is shown in Fig. 1d, where SPPs launched from the left-hand side indeed can be focused to a point on the perimeter of the dielectric cone base. In addition, the gradual change of the mode index (impedance matching) reduces the scattering loss of SPPs and the reflections from element boundaries inherent to discrete structured elements by at least one order of magnitude.

The experimental realization of the spatially varying PMMA height profile was achieved using grey-scale electron beam lithography (EBL). In this manner, the electron dose is continually varied across the sample in order to modulate the height of the PMMA layer and thus the mode index of SPPs. Fluorescence imaging [30] and leakage radiation microscopy [31] are applied to characterize the performance of the lenses (see Methods section).

We start our investigation with fluorescence imaging of SPP propagation in the Luneburg lens. SPPs at 810 nm were launched with a grating coupler 10 µm apart from the lens structure. Simultaneously, the fluorescence of the incorporated dye molecules in the PMMA excited by the SPPs was imaged on a camera to visualize the SPP propagation. Figure 2 shows the obtained fluorescence images for SPPs passing the Luneburg lens at different positions. To demonstrate that the propagation of the SPPs are modified as desired after passing the lens structure, the PMMA behind the lens was not removed and used for additional imaging. As the SPPs enter the area of the Luneburg lens (dashed circle) they are focused to a point on the perimeter of the lens on the opposite side. Even if the position of the incident SPPs is changed laterally, the SPPs are focused to the same spot as seen in Fig. 2. Because the dyes are excited by the evanescent field of the SPPs penetrating the PMMA, the local fluorescence intensity depends not only on the strength of the SPPs but also on the thickness of the PMMA at that location. Therefore, we calculated the corresponding fluorescence emission by integrating field strength of the evanescent field over the thickness of the structures to compare the ideal SPP propagation with the measured fluorescence images (lower panels in Fig. 2). The agreement between the experimental results and numerical calculations for the ideal structure is very good. The slight shadows in the measured fluorescence images are likely due to non-uniformity in the dye concentration and surface defects in the PMMA layer.

Since the visualization of the entire SPP beam path along the surface was not accessible by this technique, we used leakage radiation microscopy to image the SPP propagation on the metal-air interface. For this imaging technique, a second set of samples was fabricated on a 50 nm gold film on top of a 150 µm glass substrate. Due to the reduced thickness of the metal film SPPs from the top interface can couple through the film and generate a coherent leakage radiation into the higher index substrate [31]. For these samples no dyes were incorporated into the PMMA. Despite the difference in thickness of the gold film, no appreciable difference in the SPP propagation constant was found between the two configurations. However, imaging the leakage radiation from the back side of the substrate provided a direct quantitative analysis of the SPP propagation on the top metal surface.

Typical leakage radiation images recorded for SPPs at different wavelengths propagating through a Luneburg lens are shown in Fig. 3a-c, each exhibiting a characteristic fringe pattern arising from the interference of the directly transmitted light and the leakage radiation of the SPPs. Due to the coherence of the laser light, such interference fringes can give some information about the phase evolution of the SPPs while propagating along the surface. The phase front appears flat when launched at the grating and starts curving inside the lens leading to the focusing at a point on the perimeter, consistent with the full-wave simulation shown in Fig. 1d. The mode index of the SPPs at the centre of the Luneburg lens approaches the light cone for propagating waves at the glass substrate side (n=1.5). This leads to a decrease of the leakage radiation that can be collected by the numerical aperture of the microscope objective. Therefore, the imaged intensity inside the Luneburg lens is reduced but still visible. Figure 3d shows the corresponding height profile of the PMMA through the centre of the lens. Clearly visible are the three gold ridges for launching the SPPs at $z=0$ µm and the linear height change of the Luneburg lens.

Although the Luneburg lens was designed for a particular wavelength of 800 nm the dispersion of the PMMA around this wavelength is reasonably small. In addition, the dispersion of SPPs due to the different penetration into the dielectric medium is also not strong. All three wavelengths in Fig. 3a-3c show a clear focusing, although for longer wavelengths the focus is shifted slightly behind the lens. Such a behaviour for longer wavelengths is mainly due to the larger penetration depth of the SPPs field into the dielectric material. As a result, the overall effective mode index is smaller than what an ideal Luneburg lens requires [Eq. (1)], and hence the effective focal length is slightly longer. Nevertheless, the Luneburg lens shows a good performance within the measured 70 nm bandwidth.

To demonstrate the versatility of our method to create low-loss manipulation of SPPs solely by gradually tailoring the dielectric cladding layer, a plasmonic Eaton lens was designed and experimentally realized. The index distribution of a perfect Eaton lens that bends light 90° satisfies [32]

$$n^2 = \frac{R}{nr} + \sqrt{\left(\frac{R}{nr}\right)^2 - 1} \qquad (2)$$

Unlike the Luneburg lens, the refractive index quickly diverges as approaching the centre of the Eaton lens. Since the coating with PMMA can provide only a limited mode index range for SPPs from 1.02 up to 1.54 we are not able to realize the central part of the Eaton lens. For practical purposes we truncate the index profile at a maximum of 1.54 as shown in Fig. 4a.

A numerical simulation of the SPP propagation for the truncated Eaton lens is shown in Fig. 4b. The excited SPPs are propagating in the positive $z$-direction and bend to the right side while passing through the lens. Eventually SPPs leave the structure in the positive $x$-direction. Due to the inherent propagation loss of the SPPs the field magnitude decreases during the propagation. The truncation of the index in the centre part of the lens leads to a small deviation from the perfect 90° bend. To extract the influence of the changing field magnitude we again calculate the corresponding fluorescence intensity by taking the local thickness of the PMMA into account (Fig. 4c).

A scanning electron microscopy image of the experimentally realized structure is shown in Fig. 5 together with the corresponding cross-section of the height profile. Although the profile is more complex than the Luneburg lens the overall agreement is reasonably good; only the steepest part of the profile shows a small discrepancy from the theoretical model likely due to proximity effects during the exposure. The SPP induced fluorescence emission intensity from the structure is shown in Fig. 5c where the SPPs are propagating from the bottom in the positive *z*-direction and bend inside the Eaton lens to the right side. Figure 5d shows the simulated fluorescence intensity for the actual PMMA profile of the lens measured by AFM. Both images show similar features as the simulation done for the ideal profile (Fig. 4c).

In summary, we have experimentally demonstrated the feasibility of tailoring the propagation of SPPs by solely modifying a dielectric material on top of a metal. Since the topology of the dielectric structure is slowly varied, this technique is analogous to the well-known gradient index optics. Due to the penetration depth of the evanescent SPP field into the dielectric material the propagation constant can be spatially and gradually modified with different thicknesses of a dielectric index material. We demonstrated our approach for the examples of a plasmonic Luneburg lens and Eaton lens, both not realized at optical frequencies to date. Our approach has the potential to achieve low-loss functional plasmonic elements with a standard fabrication technology based on grey-scale electron beam lithography and is fully compatible with active plasmonics. The loss could be even further reduced by incorporating various gain materials into the dielectric material [34], leading to an increased propagation distance highly desired for all-optical devices or even plasmonic interconnects. Furthermore, this method provides a scheme to realize more complex two-dimensional plasmonic elements using transformation optics.

METHODS

The grey-scale EBL process began by determining the relationship between PMMA resist height and dose value. To enable precise control over the resist height low contrast 50k PMMA was employed, allowing a more linear resist response with respect to dose. The correlation between height and dose was then used to determine the spatially varying dose profile for the designed height/index profile. Devices for the fluorescence emission imaging are fabricated by incorporating infrared dye (IR-140, Sigma Aldrich) into the PMMA at a loading concentration of 2 mg dye to 1 mL PMMA. The dye/PMMA solution was spun and baked onto a 200-nm-thick Au film on top of a glass substrate resulting in a thickness of 500 nm. Grey-scale EBL exposure was then performed to realize the designed height profile for the Luneburg and Eaton lenses.

For our design and simulations we use as dielectric cladding layer PMMA with $\varepsilon_2 = 2.19$ on top of a gold film, whose dielectric constants are given by the Drude model $\varepsilon_m(\omega) = 10 - \frac{(1.4 \times 10^{16})^2}{\omega(\omega + 1.1 \times 10^{14} i)}$. Considering an air/dielectric/metal structure with permittivities $\varepsilon_1/\varepsilon_2/\varepsilon_m$, the dispersion relation of SPPs is implicitly given by [27]

$$\tanh(k_2 \varepsilon_2 d) = -\frac{k_1 k_2 + k_2 k_m}{k_2^2 + k_1 k_m}$$

with

$$k_{1(2,m)} = \frac{\sqrt{\beta^2 - \varepsilon_{1(2,m)} \omega^2/c^2}}{\varepsilon_{1(2,m)}} .$$

Here, $d$ is the thickness of the dielectric layer, $\beta$ represents the SPP wave vector along the propagating direction, $\omega$ is angular frequency, and $c$ denotes the speed of light in vacuum. The effective mode index of SPPs, defined as $n_{eff} = \beta/k_0$, can be obtained by numerically solving the above two equations for different heights of the dielectric layer. Once the relationship between the SPP effective index and the PMMA height is known, the PMMA profile for the plasmonic Luneburg and Eaton lenses can be readily achieved to satisfy Eqs. (1) and (2), respectively.

For the fluorescence imaging measurements, SPPs were launched by a set of gratings fabricated by focused ion beam milling 10 µm apart from the lens structure. Light from a Ti:Sa-laser at a wavelength of 810 nm was moderately focused on the grating from the backside of the sample to a spot size of 6 µm. The fluorescence emission from the IR140 dye molecules was collected with a 100x/0.95 microscope objective and imaged on a CCD camera.

For the leakage radiation microscopy, SPPs were launched by three gold ridges with a period of 805 nm and a line width of 400 nm fabricated by electron beam lithography and subsequent deposition of 50 nm gold [33]. SPPs were launched by focusing the laser light with a microscope objective (10x/0.1) on these gold ridges. The leakage radiation of the SPPs into the substrate was collected by an oil immersion objective (100x/1.3) and imaged on a CCD camera.

FIGURES

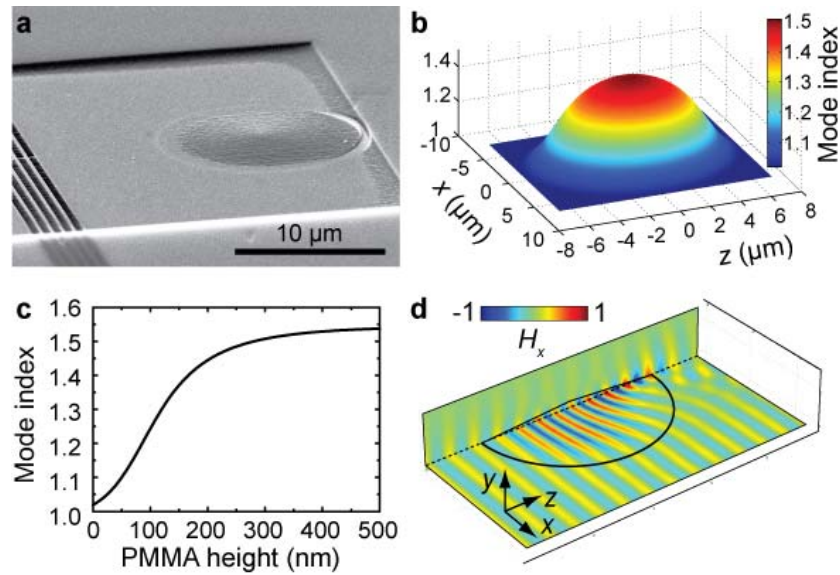

**Figure 1. Plasmonic Luneburg lens. a** Scanning electron microscopy image of a Luneburg lens on top of a gold film. The lens is made of PMMA with a diameter of 13 µm. On the left-side of the image, the SPP grating can be seen which is made of air grooves fabricated by focused ion beam milling. **b** Mode index (vertical axis) for the plasmonic Luneburg lens at a wavelength of 810 nm. **c** Relation between the PMMA height and the mode index for SPPs on a gold surface at a wavelength of 810 nm. **d** Cross-sections of the normalized magnetic field $H_x$ (at metal-dielectric interface in the x-z plane and for x=0 in the y-z plane) for SPPs propagating through the Luneburg lens in positive z-direction. The lens is designed to focus the SPPs to a point on the perimeter of the lens.

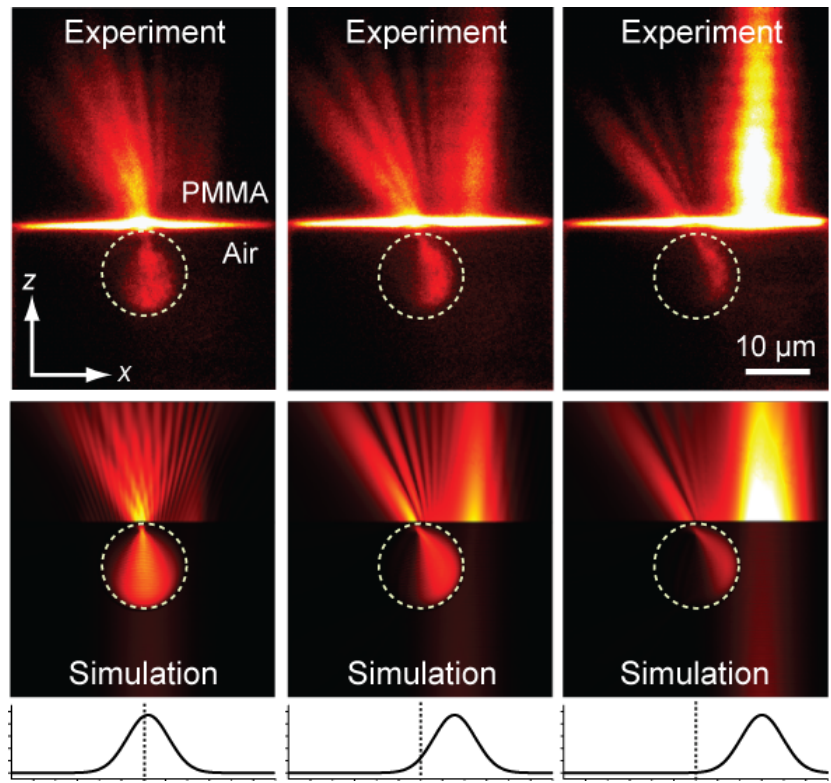

**Figure 2. Fluorescence images of the plasmonic Luneburg lens.** The upper panels show the experimentally obtained fluorescence intensity for various beam launching positions of the SPPs. The corresponding simulations are displayed in the lower panels. The fluorescence intensity is colour coded from black (low) to white (high). The position of the lenses are marked by the yellow dotted circle. The SPPs have a Gaussian intensity distribution in the $x$-direction with a full-width at half maximum of 6 µm. The images show the propagation of SPPs in positive $z$-direction for three different launching positions in the $x$-direction as indicated by the beam profile below the images (vertical dotted line marks the centre of the lens).

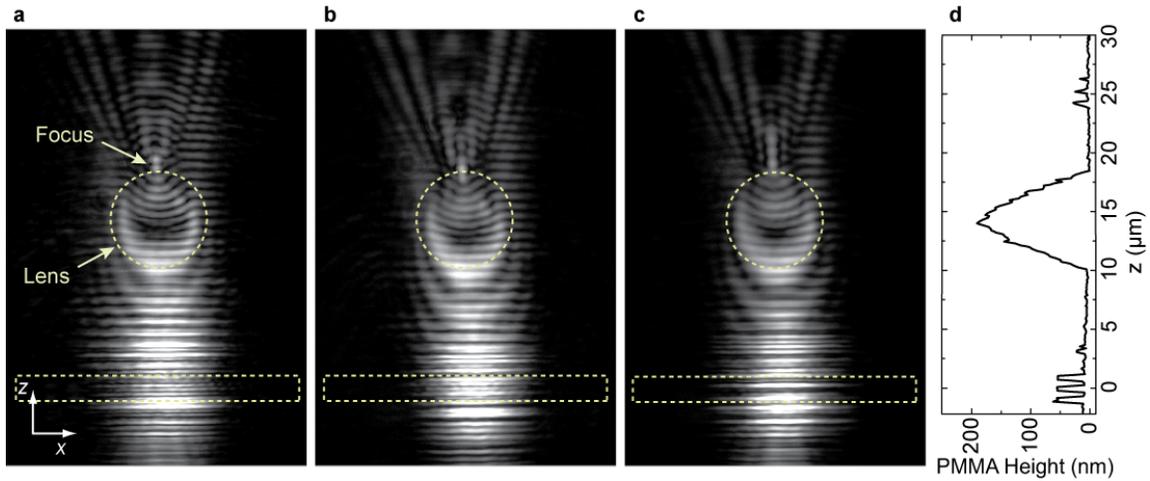

**Figure 3. Broadband performance of the plasmonic Luneburg lens.** Intensity images obtained by leakage radiation microscopy for SPPs passing a Luneburg lens for different wavelengths. **a-c** Recorded intensities for wavelengths of 770 nm (**a**), 800 nm (**b**), and 840 nm (**c**) show focusing over a 70 nm bandwidth for the plasmonic Luneburg lens. The SPPs are launched at a gold grating (area of the dashed box) 10 µm away from the lens (marked by the dashed circle). **d** Corresponding surface profile cross-section along the propagation (z) direction measured by AFM showing the height of the lens and the gratings.

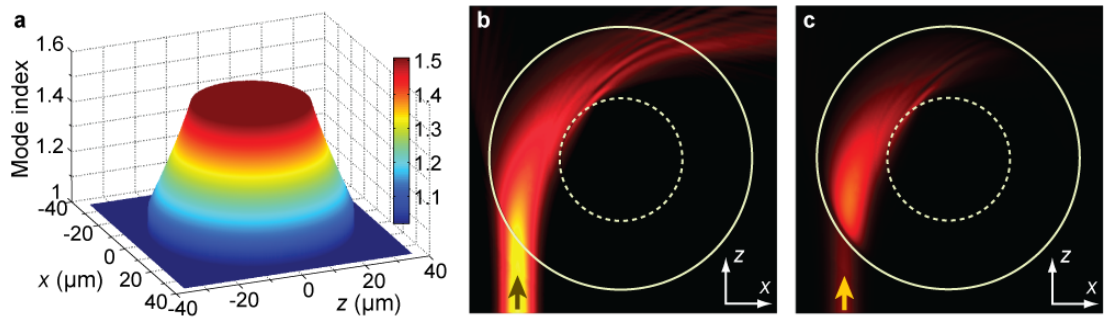

**Figure 4. Numerical simulations of the plasmonic Eaton lens. a** Truncated mode index profile for the Eaton lens with 30 µm radius. The values larger than 1.54 in the centre are cut and set to 1.54 due to the finite range provided by the index of SPPs at the metal-PMMA interface. **b** Calculated magnitude of the electric surface field for a SPP launched in positive z-direction. The solid line marks the outer diametre of the lens, and the dashed line marks the truncated index region with values larger than 1.54. The SPPs bend to the right side when passing through the lens. **c** Calculated fluorescence intensity for **b** taking the height of the dye/PMMA layer into account, visualizing the expected intensity from the structure. In the colour scale in **b** and **c** black represents low fields-amplitudes/intensities and yellow high fields-amplitudes/intensities.

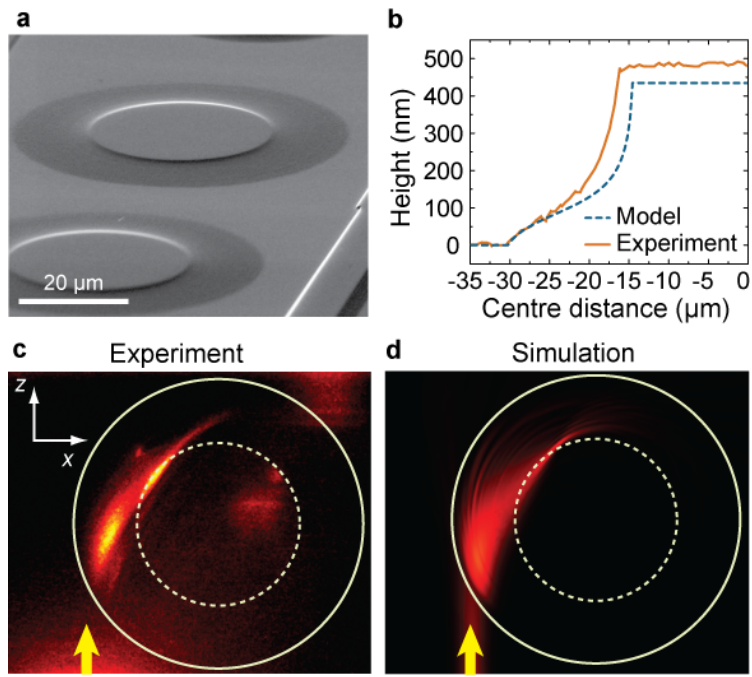

**Figure 5. Demonstration of the plasmonic Eaton lens. a** Scanning electron microscopy image of an array of Eaton lenses on top of a gold film. **b** Height profile cross-section for the left side of the lens measured by AFM (solid line) compared to the model (dashed line). **c, d** Fluorescence microscopy image and corresponding simulation of the fluorescence intensity of the Eaton lens for SPPs propagating in positive *z*-direction and bending to the right side when passing through the lens. The arrows indicate the launching position and direction of the SPPs. The solid lines mark the outer diameter of the lens and the dashed lines mark the high index region that was set to 1.54. In the colour scale black represents low and yellow high intensities.

REFERENCES

[1] Barnes, W. L., Dereux, A. & Ebbesen, T. W. Surface plasmon subwavelength optics. *Nature* **424**, 824-830 (2003).

[2] Nie, S. M. & Emery, S. R. Probing single molecules and single nanoparticles by surface-enhanced Raman scattering. *Science* **275**, 1102–1106 (1997).

[3] Fang, N., Lee, H, Sun, C. & Zhang, X. Sub-diffraction-limited optical imaging with a silver superlens. *Science* **308**, 534-537 (2005).

[4] Stockman, M. I. Nanofocusing of optical energy in tapered plasmonic waveguides. *Phys. Rev. Lett.* **93**, 137404 (2004).

[5] Srituravanich, W. et al. Flying plasmonic lens in the near field for high-speed nanolithography, *Nat. Nanotech.*, **3**, 733-737 (2008).

[6] Hohenau, A. et al. Dielectric optical elements for surface plasmons. *Opt. Lett.* **30**, 893-895 (2005).

[7] Bozhevolnyi, S. I., Volkov, V. S., Devaux, E., Laluet, J.-Y. & Ebbesen, T. W. Channel plasmon subwavelength waveguide components including interferometers and ring resonators. *Nature* 440, 508–511 (2006).

[8] Drezet, A. et al. Plasmonic crystal demultiplexer and multiports. *Nano Lett.* **7**, 1697-1700 (2007).

[9] Engheta, N. Circuits with light at nanoscales: Optical nanocircuits inspired by metamaterials. *Science* **317**, 1698-1702 (2007).

[10] Oulton, R. F. et al. Plasmon lasers at deep subwavelength scale. *Nature* **461**, 629-632 (2009).

[11] Bozhevolnyi, S. I., Erland, J., Leosson, K., Skovgaard, P. M. W. & Hvam, J. M. Waveguiding in surface plasmon polariton band gap structures. *Phys. Rev. Lett.* **86**, 3008–3011 (2001).

[12] Ditlbacher, H., Krenn, J. R., Schider, G., Leitner, A. & Aussenegg, F. R. Two-dimensional optics with surface plasmon polaritons. *Appl. Phys. Lett.* **81**, 1762-1764 (2002).

[13] Smolyaninov, I. I., Elliott, J., Zayats, A. V. & Davis, C. C. Far-field optical microscopy with a nanometer-scale resolution based on the in-plane image magnification by surface plasmon polaritons. *Phys. Rev. Lett.* **94**, 057401 (2005).

[14] Devaux, E. et al. Refractive micro-optical elements for surface plasmons: from classical to gradient index optics. *Opt. Express* **18**, 20610- 20619 (2010).

[15] Smolyaninov, I. I. Transformational optics of plasmonic metamaterials. *New Journal of Physics* **10**, 115033 (2008).

[16] Pendry, J. B., Schurig, D. & Smith, D. R. Controlling electromagnetic fields. *Science* **312**, 1780-1782 (2006).

[17] Leonhardt, U. Optical conformal mapping. *Science* **312**, 1777-1780 (2006).

[18] Schurig, D. et al. Metamaterial electromagnetic cloak at microwave frequencies. *Science* **314,** 977-980 (2006).

[19] Valentine, J., Li, J., Zentgraf, T., Bartal, G. & Zhang, X. An optical cloak made of dielectrics. *Nature Mater.* **8**, 568-571 (2009).

[20] Gabrielli, L. H., Cardenas, J., Poitras, C. B. & Lipson, M. Silicon nanostructure cloak operating at optical frequencies. *Nature Photon.* **43**, 461-463 (2009).

[21] Cheng, Q., Cui, T. J., Jiang, W. X. & Cai, B. G. An omnidirectional electromagnetic absorber made of metamaterials. *New Journal of Physics* **12**, 063006 (2010).

[22] Zentgraf, T., Valentine, J., Tapia, N., Li, J. & Zhang, X. An Optical "Janus" Device for Integrated Photonics. *Adv. Mater.* **22**, 2561–2564 (2010).

[23] Kundtz, N. & Smith, D. R. Extreme-angle broadband metamaterial lens. *Nature Mater.* **9**, 129-132 (2010).

[24] Luneburg, R. Mathematical Theory of Optics, Brown University, 1944.

[25] Eaton, J. E. On spherically symmetric lenses. *Trans. IRE Antennas Propag.* **4**, 66-71 (1952).

[26] Huidobro, P. A., Nesterov, M. L., Martín-Moreno, L. & García-Vidal, F. J. Transformation

Optics for Plasmonics. *Nano Lett.* **10**, 1985-1990 (2010).

[27]  Liu, Y., Zentgraf, T., Bartal, G. & Zhang, X. Transformational Plasmon Optics. *Nano Lett.* **10**, 1991-1997 (2010).

[28]  Renger, J. et al. Hidden progress: broadband plasmonic invisibility. *Opt. Exp.* **18**, 15757-15768 (2010).

[29]  Aubry, A. et al. Plasmonic Light-Harvesting Devices over the Whole Visible Spectrum. *Nano Lett.* **10**, 2574 (2010).

[30]  Ditlbacher, H. et al. Fluorescence imaging of surface plasmon fields. *Appl. Phys. Lett.* **80**, 404-406 (2002).

[31]  Drezet, A. et al. Leakage radiation microscopy of surface plasmon polaritons. *Materials Science and Engineering B* **149**, 220–229 (2008).

[32]  Danner, A. J. & Leonhardt, U., 2009 Conference on Lasers and Electro-Optics (CLEO), Baltimore, MD, USA.

[33]  Radko, I. P. et al. Efficiency of local surface plasmon polariton excitation on ridges. *Phys. Rev. B* **78**, 115115 (2008).

[34]  De Leon, I. & Berini, P. Amplification of long-range surface plasmons by a dipolar gain medium. *Nature Photon.* **4**, 382 - 387 (2010).

**Acknowledgements**

We acknowledge funding support from the US Army Research Office (ARO) MURI program W911NF-09-1-0539 and US NSF Nano-scale Science and Engineering Center CMMI-0751621.

**Author contributions**

T. Z., Y. L., and J. V. conceived and designed the experiments. T. Z. and M. H. M. performed the experiments and analyzed the data. Y. L. designed the structures and performed the numerical simulations. J. V. and M. H. M. fabricated the samples. X. Z. guided the theoretical and experimental work. All authors discussed the results and co-wrote the manuscript.

**Additional information**

Reprints and permission information is available online at http://npg.nature.com/reprintsandpermissions/.

Correspondence and requests for materials should be addressed to X. Z.